\documentclass[epj]{svjour}
\usepackage{latexsym}
\usepackage[dvips,dvipdf]{graphicx}
\usepackage{epsf}
\begin{document}
\title{Solving Bethe-Salpeter equation in Minkowski space
}
\author{V.A. Karmanov\inst{1} \and J. Carbonell\inst{2}
}                     
%
%
\institute{Lebedev Physical Institute, Leninsky Prospekt 53, 119991
Moscow, Russia \and Laboratoire de Physique Subatomique et Cosmologie,
 53 avenue des Martyrs,
38026 Grenoble, France}
\date{Received: date / Revised version: date}
%
\abstract{We develop a new method of solving Bethe-Salpeter (BS)
equation in Minkowski space. It is based on projecting the BS
equation on the light-front (LF) plane and on the Nakanishi
integral representation of the BS amplitude. This method is valid
for any kernel given by the irreducible Feynman graphs. For
massless ladder exchange, our approach reproduces analytically the
Wick-Cutkosky equation. For massive ladder exchange, the numerical
results coincide with the ones obtained by Wick rotation.
} 
%
\maketitle
%

\section{Introduction}\label{intr}

BS equation \cite{SB_PR84_51}  is an
important tool to study the relativistic  bound state problem in a field
theory framework (see for review~\cite{nakanishi}).
For a bound state of total momentum $p$ and in case of equal mass particles, it reads
\begin{eqnarray}\label{bs}
\Phi(k,p)&=&\frac{i^2}{\left[(\frac{p}{2}+k)^2-m^2+i\epsilon\right]
\left[(\frac{p}{2}-k)^2-m^2+i\epsilon\right]}
\nonumber\\
&\times&\int \frac{d^4k'}{(2\pi)^4}iK(k,k',p)\Phi(k',p),
\end{eqnarray}
where $\Phi$ is the BS amplitude, $iK$ the interaction kernel, $m$
the mass of the constituents and $k$ their relative momentum. We
will denote by $M=\sqrt{p^2}$ the total mass of the bound state,
and by $B=2m-M$ its binding energy.

It was recognized from the very beginning that, when formulated in
Minkowski space, the  BS equation has singularities which make difficult
to find its solution. These singularities are due to the free propagators
of the constituent particles
\begin{eqnarray}\label{G0}
&&G_0^{(12)}(k,p) =G_0^{(1)}G_0^{(2)}=
\nonumber\\
&&=\frac{i}{(\frac{p}{2}+ k)^2-m^2+i\epsilon} \;
\frac{i}{(\frac{p}{2}-k)^2-m^2+i\epsilon}
\end{eqnarray}
but can also result from the interaction kernel itself.

To overcome this difficulty, Wick \cite{WICK_54}  formulated the BS
equation in the Euclidean space, by  rotating the relative energy in the
complex plane  $k_0\to ik_0$. This "Wick rotation" led to a well defined
integral equation which can be solved by standard methods. Most of
practical applications of the BS equation have been achieved using this
technique  \cite{nakanishi} and recent developments make its solution a
trivial numerical task  \cite{NT_FBS_96}. Another method -- the
variational approach in the configuration Euclidean space -- was recently
developed in~\cite{efimov}. Whereas the total mass of the system is
unchanged by the Wick rotation, the original BS amplitude is however lost
and the "rotated" one can no longer be used in calculating other physical
observables, like for instance form factors.

Thus, fifty years after its formulation, obtaining the BS solutions in the
Minkowski space is still a field of active research. A successful attempt
was presented in \cite{KW}, based on the Nakanishi integral representation of the BS
function \cite{nak63}. However, formal developments displayed in \cite{KW}
are a matter of art and the obtained equation has been
derived and solved only for the ladder kernel.
Another approach in Minkowski space for separable interactions was developed
in~\cite{bbmst} and applied to the nucleon-nucleon system.
On another hand, an equation obtained by projecting the original BS
equation on the LF plane, was derived and solved in \cite{sfcs}.
An approximate LF kernel was there obtained as an expansion of the BS one
but the original BS amplitude has not been reconstructed from its LF projection.

The aim of this paper and the forthcoming one is to present a new method
of solving the BS equation without using the Wick rotation. Our method is based on
an integral transform of the initial equation which removes the
singularities of the BS amplitude. This integral transform
consists in projecting the BS equation on the LF plane, defined by
$\omega\cdot x=0$ with $\omega^2=0$ \cite{cdkm}. The particular choice
$\omega=(\omega_0,\vec{\omega})=(1,0,0,-1)$ results in the standard LF
form $t+z=0$ and in the LF projection used in \cite{sfcs}.
In our approach, the BS amplitude
maps onto the LF wave function while the transformed equation -- in contrast to
\cite{sfcs} -- is derived  without any approximation.
This equation remains equivalent to the original BS one, therefore providing
the same binding energies,
and the initial BS amplitude is easily reconstructed from its solution.
Although results presented here concern only the ladder
kernel, our method is not restricted to a particular interaction.
For more complicated kernels, {\it e.g.} the cross box, calculations become more
lengthy, but the additional difficulties are due
to evaluating the Feynman diagram itself and not to the solution of the equation.

In order to present the method more distinctly, we consider the
case of zero total angular momentum and spinless particles.

The plan of the paper is the following. In sect. \ref{project}, we
give the integral transform used to project the BS equation on the
LF plane and we derive a new and equivalent equation. In sect.
\ref{le}, the corresponding ladder exchange kernel is calculated
analytically. In sect. \ref{num}, the numerical solutions for the
ladder case are found and compared  to the results obtained using
other methods in Euclidean space. Sect. \ref{concl} contains
concluding remarks. Details of the calculations are given in
appendices \ref{deriv}, \ref{LFeq} and \ref{calcI}. The results
concerning cross box kernels are presented in the next paper
\cite{ckII}.

\section{Projecting the BS equation on the LF plane}\label{project}

Our method is inspired by an existing relation between the BS
amplitude $\Phi(k,p)$ and the two-body LF wave function
$\psi(\vec{k}_{\perp},x)$. This wave function  can be obtained by
projecting the BS amplitude on the LF plane. We will apply below
the LF projection to the BS equation in Minkowski space. Though
this projection can be considered as a formal transform, we will
start by reviewing its derivation, in order to show more clearly
how the singular behaviour of $\Phi(k,p)$ gives rise to a
non-singular $\psi(\vec{k}_{\perp},x)$.

BS amplitude is defined as the matrix element between the vacuum $\langle
0|$ and a state $|p\rangle$ of the time ordered product of two
Heisenberg operators:
\begin{equation}\label{BSA}
\Phi(x_1,x_2,p)=\langle 0|T\left\{\varphi(x_1)\varphi(x_2)\right\}|p\rangle.
\end{equation}
In general, the state vector $|p\rangle$ can be taken in different
representations. In the LF quantization, it has the form:
(see {\it e.g.} eq. (3.1) from \cite{cdkm}):
\begin{eqnarray}
\label{state}
|p\rangle &=& \int\;\psi(k_1,k_2,p,\omega\tau) 2(\omega\cdot p)
\delta(k_1+k_2-p-\omega\tau)d\tau
\nonumber\\
&\times& (2\pi)^{3/2}\;
 \frac{d^3k_1}{(2\pi)^{3/2}\sqrt{2\varepsilon_{k_1}}}
 \frac{d^3k_2}{(2\pi)^{3/2}\sqrt{2\varepsilon_{k_2}}}\;
a^{\dagger}_{\vec{k}_1}a^{\dagger}_{\vec{k}_2}|0\rangle
\nonumber\\
&+&\; \ldots\;,
\end{eqnarray}
where $a^{\dagger}_{\vec{k}}$ is the creation operator and
$\varepsilon_{k}=\sqrt{m^2+\vec{k}^2}$. The two-body Fock
component $\psi$ is shown explicitly, whereas the higher ones are
implied. All the four-momenta are on the corresponding mass shells
$k_i^2=m^2$, $p^2=M^2$, $(\omega\tau)^2=0$ and fulfill the
conservation law
 $$
k_1+k_2=p+\omega\tau. $$

Projecting the BS amplitude $\Phi(x_1,x_2,p)$ on LF plane means that its
arguments are constrained to $\omega\cdot x_1= \omega\cdot x_2=0$. Coming
to the momentum space, we still keep this constrain. Let us evaluate the
quantity:
\begin{eqnarray}
\label{J}
J(k_1,k_2,p)&\equiv& \int d^4x_1 d^4x_2\delta(\omega\cdot x_1) \;\delta(\omega\cdot x_2)
e^{i (k_1\cdot x_1+k_2\cdot x_2)}
\nonumber\\
&&\times {\Phi}(x_1,x_2,p).
\end{eqnarray}
We substitute here the right-hand side of (\ref{BSA}) with $|p\rangle$
given by (\ref{state}). On the LF plane the Heisenberg field $\varphi(x)$
in (\ref{BSA}) turns into the Schr\"odinger (free) one, represented as
\[\varphi(x)= \int\left(a^{\dagger}_{\vec k}e^{-ik\cdot x}+a_{\vec
k}e^{ik\cdot x}\right)
\frac{d^3k}{(2\pi)^{3/2}\sqrt{2\varepsilon_{k}}}.\] Then the
two-body component $\psi$ only  survives in $|p\rangle$ and
$J(k_1,k_2,p)$ is expressed through it.

Now express $\Phi(x_1,x_2,p)$ in (\ref{J}) through its Fourier transform.
Translational invariance imposes $\Phi$ to have the  form
 $$
\Phi(x_1,x_2,p)=\frac{1}{(2\pi)^{3/2}} \; \tilde{\Phi}(x,p) \;e^{-ip
\cdot(x_1+x_2)/2},
 $$
where $\tilde{\Phi}(x,p)$ is the reduced amplitude
and $x=x_1-x_2$. It is expressed through the momentum space BS amplitude:
 $$ \tilde\Phi(x,p)= \int \frac{d^4k}{(2\pi)^4}
\;\Phi(k,p)\; e^{-ik\cdot x},
 $$
where $\Phi(k,p)$ satisfies the BS equation (\ref{bs}). Substituting
theses formulas in (\ref{J}), we find that $J(k_1,k_2,p)$ is expressed
through the integral $\int_{-\infty}^{\infty}\Phi(k+\beta\omega,p)d\beta$.
Comparing two expressions, we obtain the relation~\cite{cdkm}:
\begin{equation}\label{lfwf}
\psi(\vec{k}_{\perp},x)=\frac{(\omega\cdot k_1)(\omega\cdot k_2)}{\pi(\omega\cdot p)}
\int_{-\infty}^{\infty}\Phi(k+\beta\omega,p)d\beta.
\end{equation}
In the standard LF approach the $\beta$-integration in (\ref{lfwf}) turns
into the $k_-$-integration with $k_-=k_0-k_z$. Wave function
$\psi(\vec{k}_{\perp},x)$ in (\ref{lfwf}) is parametrized in terms of the
standard LF variables $\vec{k}_{\perp},x$ (see \cite{cdkm}):
\begin{equation}\label{lfvar}
\vec{k}_{\perp}=\vec{k}_{1\perp}-x\vec{p}_{\perp},
  \qquad
x=\frac{\omega\cdot k_1}{\omega\cdot p}.
\end{equation}
The $\perp$-components are orthogonal to $\vec{\omega}$.

LF wave function $\psi$, as any wave function, has no
singularities in physical domain. Equation  (\ref{lfwf}) can be
thus viewed as an integral transformation of the BS amplitude
leading to a non-singular function. It suggests to apply this
transformation to the BS equation itself:
\begin{eqnarray}\label{BST}
&&\int d\beta\Phi(k+\beta\omega,p)=
\\
&&\int d\beta G_0^{(12)}(k+\beta\omega,p)
\int\frac{d^4k'}{(2\pi)^4} iK(k+\beta\omega,k',p)\Phi(k',p),
\nonumber
\end{eqnarray}
in order to obtain an  equivalent equation free of singularities. This
constitutes the key point of this work.

Apart from the trivial kinematical factor $\frac{(\omega\cdot
k_1)(\omega\cdot k_2)}{\pi(\omega\cdot p)}$, the left-hand side of
(\ref{BST}) is the LF wave function $\psi$, eq. (\ref{lfwf}),
where\-as, the right-hand side still contains the "non-projected"
BS amplitude $\Phi(k',p)$. To make the LF wave function appear
explicitly in the right-hand side too, and thus formulate an
equation in terms of $\psi$, would need to invert equation
(\ref{lfwf}). Instead, we substitute, in both sides of the
equation (\ref{BST}), the BS amplitude in terms of the Nakanishi
integral representation \cite{nakanishi,nak63}:
\begin{eqnarray}\label{bsint}
\Phi(k,p)&=&\frac{-i}{\sqrt{4\pi}}\int_{-1}^1dz'\int_0^{\infty}d\gamma'
\\
&\times&
\frac{g(\gamma',z')}{\left[\gamma'+m^2
-\frac{1}{4}M^2-k^2-p\cdot k\; z'-i\epsilon\right]^3}.
\nonumber
\end{eqnarray}
 In more general form of this representation, the
denominator appears in the degree $2+n$, where $n$ is a dummy
integer parameter. For simplicity, we chose here its minimal value
$n=1$. Greater value of $n$ may result in a more smooth solution
\cite{KW}.

A similar representation exists for non-zero angular momentum. It
is valid for rather wide class of the solutions which are
consistent with the perturbation-theoretical analyticity. This
leads (see appendix \ref{deriv} for the detail of calculations) to
the following equation for the weight function $g(\gamma,z)$:
\begin{eqnarray} \label{bsnew}
&&\int_0^{\infty}\frac{g(\gamma',z)d\gamma'}{\Bigl[\gamma'+\gamma +z^2 m^2+(1-z^2)\kappa^2\Bigr]^2}
=
\nonumber\\
&&\int_0^{\infty}d\gamma'\int_{-1}^{1}dz'\;V(\gamma,z;\gamma',z') g(\gamma',z'),
\end{eqnarray}
This is just the eigenvalue equation of our method. It is equivalent to
the initial BS equation (\ref{bs}).
The total mass $M$ of the system appears on both sides
of equation (\ref{bsnew}) and is contained in the parameter
\begin{equation}\label{kappa2}
\kappa^2 = m^2- \frac{1}{4}M^2.
\end{equation}
 As calculations \cite{KW} show,  $g(\gamma,z)$ may be
zero in an interval $0\le \gamma \le \gamma_0$. The exact value
where it differs from zero is determined by the equation
(\ref{bsnew}) itself.

The kernel $V$, appearing in the right-hand side of eq. (\ref{bsnew}), is
related to the kernel $iK$ from the BS equation by
\begin{eqnarray}\label{V}
V(\gamma,z;\gamma',z')&=&
\frac{\omega\cdot p}{\pi}\int_{-\infty}^{\infty}\frac{-iI(k+\beta
\omega,p)d\beta}
{\left[(\frac{p}{2}+k+\beta\omega)^2-m^2+i\epsilon\right]}
\nonumber\\
&\times&\frac{1}{\left[(\frac{p}{2}-k-\beta\omega)^2-m^2+i\epsilon\right]},
\end{eqnarray}
with
\begin{equation}\label{I}
I(k,p)=\int \frac{d^4k'}{(2\pi)^4}\frac{iK(k,k',p)}
{\left[{k'}^2+p\cdot k' z'-\gamma'-\kappa^2+i\epsilon\right]^3}.
\end{equation}

The singularities in the BS equation are removed by the analytical
integration over $\beta$. Equation (\ref{bsnew}) is valid for an arbitrary
kernel $iK$, given by a Feynman graph. The particular cases of the ladder
kernel and of the Wick-Cutkosky model \cite{WICK_54,CUTKOSKY_PR96_54} are
detailed in the next section and for the cross ladder kernel -- in the
next paper \cite{ckII}. Once $g(\gamma,z)$ is known, the BS amplitude can
be restored by eq. (\ref{bsint}).

The variables ($\gamma,z$) are related to the standard LF variables
(\ref{lfvar}) as $\gamma=k_{\perp}^2$, $z=1-2x$. The LF wave function can
be easily obtained by
\begin{equation} \label{lfwf3a}
\psi(k_\perp,x)=\frac{1}{\sqrt{4\pi}}\int_0^{\infty}\frac{x(1-x)g(\gamma',1-2x)d\gamma'}
{\Bigl[\gamma'+k_\perp^2 +m^2-x(1-x)M^2\Bigr]^2}.
\end{equation}

Eq. (\ref{bsnew}) can be transformed to the equation for the LF wave
function $\psi(k_\perp,x)$ (eq. (\ref{psi}) in appendix \ref{deriv}),
though this requires inverting the kernel in the left-hand side of
(\ref{bsnew}). The initial BS equation (\ref{BST}), projected on the LF
plane, can be also approximately transformed (see appendix \ref{LFeq}) to
the LF equation:
\begin{eqnarray}\label{eq1}
&&\left(\frac{\vec{k}^2_{\perp}+m^2}{x(1-x)}-M^2\right)
\psi(\vec{k}_{\perp},x)=
\\
&&-\frac{m^2}{2\pi^3}\int\psi(\vec{k}'_{\perp},x')
V_{LF}(\vec{k}'_{\perp},x';\vec{k}_{\perp},x,M^2)
\frac{d^2k'_{\perp}dx'}{2x'(1-x')}
\nonumber
\end{eqnarray}
with the LF kernel $V_{LF}$ given, for ladder exchange, by eq.
(\ref{lfdlad}) in appendix \ref{LFeq}.

It is worth noticing that the LF wave function (\ref{lfwf3a}) is different
from the one obtained by solving the ladder LF equation (\ref{eq1}), as
it was done e.g. in ref. \cite{mariane}. The physical reason lies in the fact that
the iterations of the ladder BS kernel (Feynman graph) and the ladder LF
kernel (time-ordered graphs) generate different intermediate states.
The LF kernel and its iterations contain in the intermediate state only one
exchanged particle,
whereas the iterations of the ladder Feynman kernel contain also,
many-body states with increasing number of exchanged particles (stretched boxes).
This leads to a difference in the binding energies, which is however small
\cite{mariane}.
Formally, this difference arises because of the approximations --
explained in appendix \ref{LFeq} -- which are made in  deriving  eq. (\ref{eq1})
from (\ref{BST}).
However, for a kernel given by a finite
set of irreducible graphs,  both BS  (\ref{bs})
and LF (\ref{eq1}) equations are already approximate and it is not evident
which of them is more "physical".
The physically transparent interpretation of the LF
wave function  makes it often more attractive.

\section{Ladder kernel}\label{le}

We calculate here the kernel $V(\gamma,z;\gamma',z')$ of equation
(\ref{bsnew}) for the ladder BS kernel, which  reads:
\begin{equation}
\label{ladder}
iK^{(L)}(k,k',p)=\frac{i(-ig)^2}{(k-k')^2-\mu^2+i\epsilon}.
\end{equation}
We substitute it in eq. (\ref{I}), then substitute (\ref{I}) in (\ref{V})
and calculate the integrals. The details of these calculations are given
in appendix \ref{calcI}. The result reads:
\begin{equation} \label{Kn}
V^{(L)}(\gamma,z;\gamma',z')=\left\{
\begin{array}{ll}
W(\gamma,z;\gamma',z'),&\mbox{if $-1\le z'\le z\le 1$}\\
W(\gamma,-z;\gamma',-  z'),&\mbox{if $-1\le z\le z'\le 1$}
\end{array}\right.
\end{equation}
where $W$ has the form:
\begin{eqnarray}
W(\gamma,z;\gamma',z') =
\frac{\alpha m^2}{2\pi}
 \frac{(1-z)^2}{\gamma+z^2m^2+(1-z^2)\kappa^2}&&
\nonumber\\
\times
\frac{1}{b_2^2(b_+ -b_-)^3}
\left[
\frac{(b_+ -b_-)(2b_+ b_- -b_+ -b_-)}{(1-b_+)(1-b_-)}\right.&&
\nonumber\\
\left.
+2b_+ b_- \log \frac{b_+ (1-b_-)}{b_- (1-b_+)}\right]&&
\label{W}
\end{eqnarray}
with $\alpha=g^2/(16\pi m^2)$ and
\begin{eqnarray*}
b_\pm &=& -\frac{1}{2b_2} \;\left( b_1 \pm \sqrt{b_1^2-4b_0b_2}\right),
\\
b_0 &=& (1-z)\mu^2,\\
b_1 &=& \gamma+\gamma' - (1-z)\mu^2 - \gamma' z -\gamma z'
\\
&+&(1-z')\left[z^2m^2+(1-z^2)\kappa^2\right], \cr
b_2 &=& -\gamma (1-z') \\
&-&(z-z') \left[  (1-z)(1-z')\kappa^2
+(z+z'-zz') m^2 \right].
\end{eqnarray*}

In the case $\mu=0$ (which constitutes the original Wick-Cutkosky model
\cite{WICK_54,CUTKOSKY_PR96_54}) we get, in particular, $b_0=b_-=0$ and
eq. (\ref{W}) obtains a more simple analytical expression that gives for
the kernel:
\begin{eqnarray}\label{KK}
&&V^{(L)}(\gamma,z;\gamma',z')=\frac{\alpha m^2}{2\pi}
\frac{1}{\Bigl[\gamma+z^2m^2+(1-z^2)\kappa^2 \Bigr]}
\nonumber\\
&&\phantom{V^{(L)}(\gamma,z;\gamma',z')=}\times
\frac{1}{\Bigl[\gamma'+{z'}^2m^2+(1-{z'}^2)\kappa^2 \Bigr]}
\nonumber\\
&\times&\left\{
\frac{\theta(z-z')}{\Bigl[\gamma +\gamma'\frac{(1-z)}{(1-z')}
+z^2m^2+(1-z^2)\kappa^2\Bigr]}
\frac{(1-z)}{(1-z')}\right.
\nonumber\\
&+&\left.
\frac{\theta(z'-z)}{\Bigl[\gamma +\gamma'\frac{(1+z)}{(1+z')}
+z^2m^2+(1-z^2)\kappa^2\Bigr]}
\frac{(1+z)}{(1+z')}\right\}.
\nonumber\\
&&
\end{eqnarray}

We search for a solution of (\ref{bsnew}) in the form:
\begin{equation}\label{g0}
g(\gamma,z)=\delta(\gamma)\; g(z).
\end{equation}
The integration over $\gamma'$ in both sides of equation (\ref{bsnew})
drops out. By setting $\gamma'=0$ everywhere, the kernel (\ref{KK}) turns into:
\begin{eqnarray*}
V^{(L)}(\gamma,z;\gamma'=0,z')=
\frac{1}{\Bigl[\gamma +z^2m^2+(1-z^2)\kappa^2\Bigr]^2}&& \\
\times
\frac{\alpha}{2\pi}
\frac{m^2}{\Bigl[{z'}^2m^2+(1-{z'}^2)\kappa^2\Bigr]}&&\\
\times
\left[
\frac{(1-z)}{(1-z')}\theta(z-z')
+
\frac{(1+z)}{(1+z')}\theta(z'-z)
\right].&&
\end{eqnarray*}
The prefactor
\[\frac{1}{\Bigl[\gamma +z^2m^2+(1-z^2)\kappa^2\Bigr]^2}\]
is the same in both sides of equation (\ref{bsnew}) and cancels.
The $\gamma$-dependence disappears thus from the equation which takes the simplified form:
\begin{equation}
\label{bsnew0}
g(z)=\frac{\alpha}{2\pi}\int_{-1}^{1}dz'\;\tilde{V}(z,z')g(z')
\end{equation}
with
\begin{eqnarray}\label{Kt}
\tilde{V}(z,z')&=&\frac{m^2}{m^2-\frac{1}{4}(1-{z'}^2)M^2}
\nonumber\\
&\times&\left\{
\begin{array}{ll}
\frac{(1-z)}{(1-z')},&\quad\mbox{if $-1\le z'\le z\le 1$}\\
\frac{(1+z)}{(1+z')},&\quad\mbox{if $-1\le z\le z' \le 1$}
\end{array}
\right.
\end{eqnarray}
It exactly coincides with the Wick-Cutkosky equation
\cite{nakanishi,WICK_54,CUTKOSKY_PR96_54}.

Notice that in the $\mu=0$ case the LF wave function (\ref{lfwf3a})
obtains the simple form~\cite{karm80,Saw}:
\begin{equation}\label{lfwf0}
\psi(k_\perp,x)=\frac{x(1-x)g(1-2x)}
{\sqrt{4\pi}\Bigl[k_\perp^2 +m^2-x(1-x)M^2\Bigr]^2}
\end{equation}
with a known analytic dependence on $k_\perp$ variable.

\section{Numerical results}\label{num}

Equation (\ref{bsnew}) with the ladder kernel (\ref{Kn}) has been solved
by using the same method as in \cite{mariane}, {\it i.e.} by expanding the
solution $g$ on a spline basis \cite{PAYNE87}
\begin{equation}\label{g_spline}
 g(\gamma,z) =\sum_{ij} g_{ij} S_{i}(\gamma)S_{j}(z)
\end{equation}
over a compact integration domain $\Omega=[0,\gamma_{max}]\times[-1,+1]$
and validating the equation at some well chosen ensemble of collocation
points $\{\bar\gamma_{i},\bar{z}_j\}\subset\Omega$. The unknown
coefficient $g_{ij}$ are determined by solving the resulting generalized
eigenvalue matrix equation
\begin{equation}\label{LS}
\lambda \; B(M) g= A(M)g
\end{equation}
in which matrices $B$ and $A$  represent respectively the integral
operators of the left- and right-hand sides of (\ref{bsnew}). They both
depend on the total mass of the system $M$ through the parameter $\kappa$
defined in (\ref{kappa2})
and the solution of the equation is  provided by the values of $M$ such
that $\lambda(M)=1$. For the ladder kernel, the coupling constant $\alpha$
appears linearly in $A$ and the problem can be formulated equivalently
\[ {1\over\alpha} \; B(M) \;g= {A(M)\over\alpha}\; g\]
in which the inverse of the coupling constant appears as the eigenvalue
of a linear system parametrised by $M$.

\begin{figure}[ht!]
\begin{center}
\includegraphics[width=8cm]{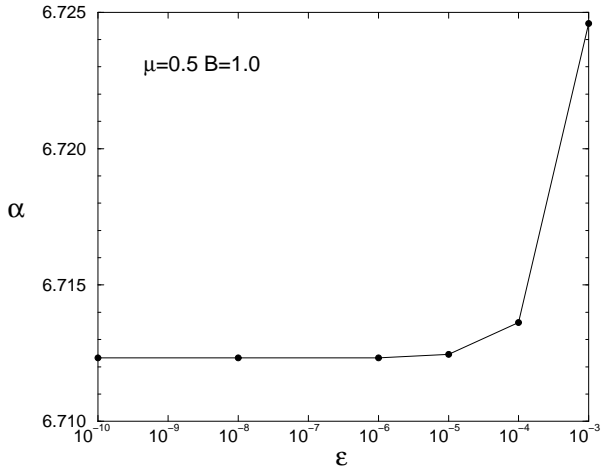}
\end{center}
\caption{Dependence of the coupling constant on the $\varepsilon$
parameter of (\ref{reg}) for $\mu=0.5$ and  $B=1.0$.}
\label{alpha_epsilon}
\end{figure}

It turns out that the discretized integral operator $B$ has very small
eigenvalues. They are unphysical but  make unstable the solution of the
system (\ref{LS}). To regularize $B$, we have added a small constant
$\varepsilon$ to its diagonal part~\cite{Korobov} on the form:
\begin{equation} \label{reg}
B_{ij}\to B_{ij} +\varepsilon \;N_{ij}
\end{equation}
where $N_{ij}$ is the equivalent of the Kronecker symbol
$\delta_{ij}$ in the bidimensional spline  basis. This procedure
allows us to obtain stable eigenvalues with an accuracy of the
same order than $\varepsilon$ until values of $\varepsilon$ as
small as $10^{-12}$. We have plotted in fig. \ref{alpha_epsilon}
the dependence of the coupling constant on  $\varepsilon$ for a
system with binding energy $B=1.0$ and $\mu=0.5$. The convergence
is very fast and a value $\varepsilon=10^{-4}$ is enough to ensure
a 4 digits stability on $\alpha$. The real accuracy of a
calculation is actually not determined by $\varepsilon$ but rather
by the grid parameters which were kept fixed in the results of
fig. \ref{alpha_epsilon}. These are essentially the value
$\gamma_{max}$ and the number of intervals $N_\gamma$ and $N_z$ in
each direction of $\Omega$.

\begin{table}[ht!]
\begin{center}
\caption{Coupling constant values
as a function of the binding energy for  $\mu=0.15$ and $\mu=0.5$
obtained with $\gamma_{max}=3$, $N_{\gamma}=12$, $N_z=10$ and
$\varepsilon=  10^{-6}$.}
\label{tab1}       
\begin{tabular}{ccc}
\hline\noalign{\smallskip}
$B$ & $\alpha(\mu=0.15)$ &  $\alpha(\mu=0.50)$  \\
\noalign{\smallskip}\hline\noalign{\smallskip}
0.01   &   0.5716           & 1.440 \\
0.10   &   1.437            & 2.498 \\
0.20   &   2.100            & 3.251 \\
0.50   &   3.611            & 4.901 \\
1.00   &   5.315            & 6.712 \\
\noalign{\smallskip}\hline
\end{tabular}
\end{center}
\end{table}

By keeping $\varepsilon=10^{-6}$ fixed and varying the grid
parameters to ensure four digits accuracy, we obtain for
$\mu=0.15$ and $\mu=0.5$ the values displayed  in  table
\ref{tab1}. They correspond to $\gamma_{max}=3$, $N_{\gamma}=12$,
$N_z=10$. With all shown digits, they are in full agreement with
the results we have obtained, similarly to \cite{mariane}, by
using the Wick rotation and the method of \cite{NT_FBS_96}.
Increasing $\varepsilon$ to $10^{-4}$ changes at most one unit in
the last digit. This demonstrates the validity of our approach.

\begin{figure*}[ht!]
\centering
\includegraphics[width=8cm]{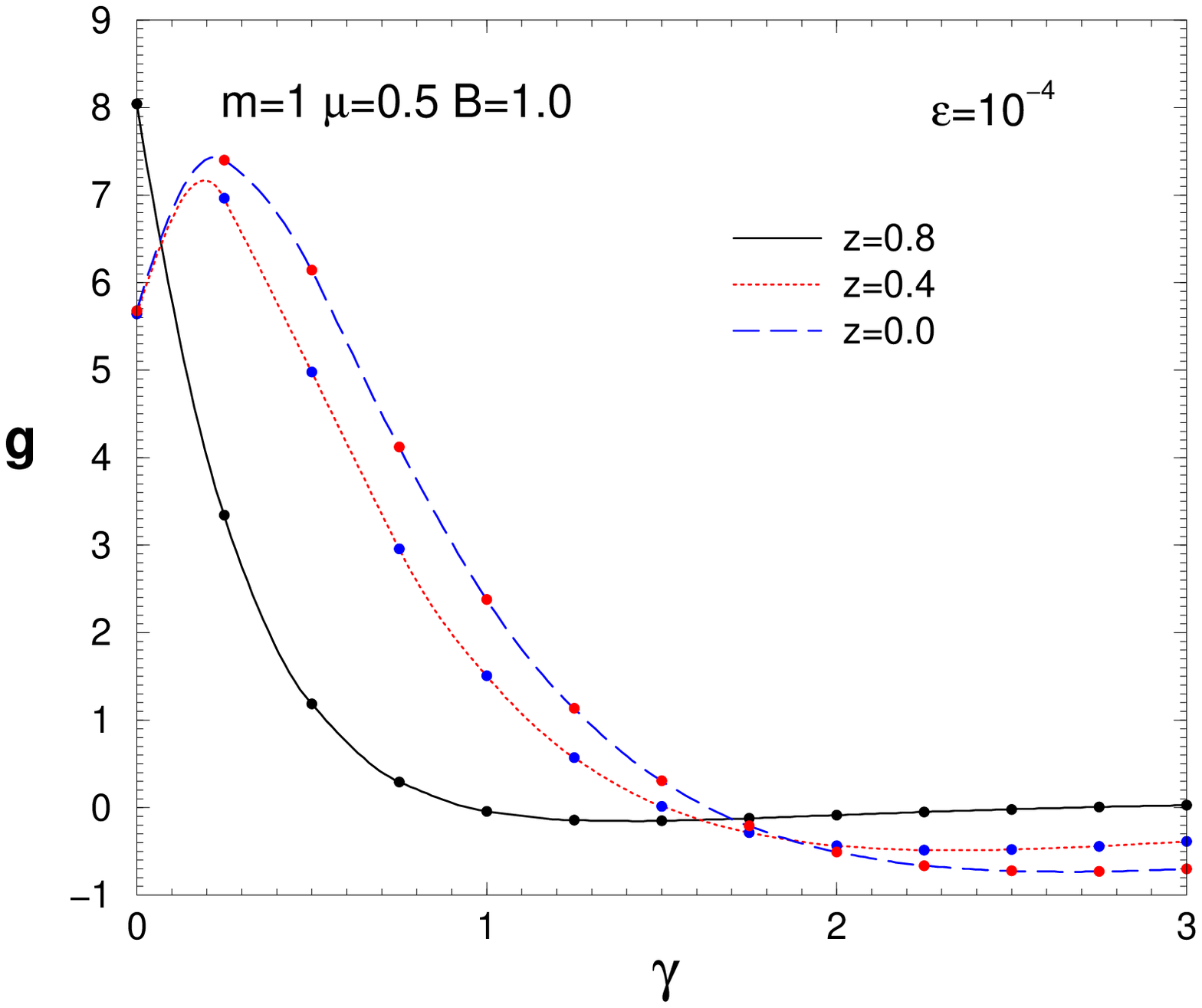}
\hspace{0.5cm}
\includegraphics[width=8cm]{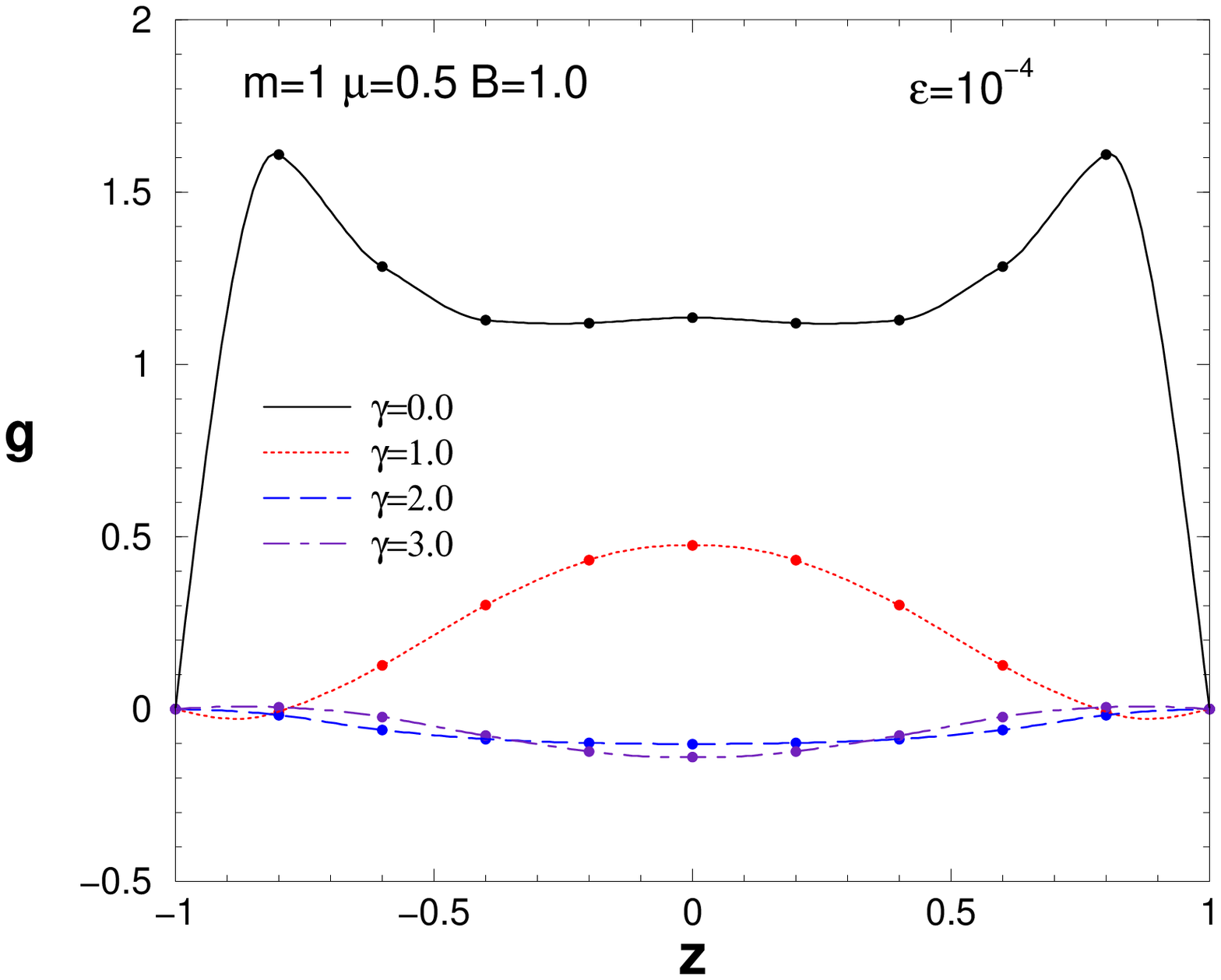}
\caption{Function $g(\gamma,z)$ for $\mu=0.5$ and  $B=1.0$.
On left -- versus $\gamma$ for fixed values of $z$ and
on right -- versus $z$ for a few fixed values of $\gamma$.}
\label{g_k}
\end{figure*}

We would like to remark the  striking stability of the results with
respect to $N_{\gamma}$ the number of grid points on $\gamma$. The value
$N_{\gamma}=12$ used in our calculations was only for drawing purposes. In
fact, the accuracy in calculating the eigenvalues in table  \ref{tab1} is
reached with $N_{\gamma}=1$. This means that on the practical point of
view our method leads to an  equation whose solution is mostly
one-dimensional and a number of grid points of the order of 10 on
$z$-variable  is enough to ensure an  accuracy better than  $10^{-4}$.

The  weight function $g$ for a system with $\mu=0.5$ and  $B=1.0$
is plotted in fig. \ref{g_k}. It has been obtained with
$\varepsilon=10^{-4}$ and the same grid parameters than in table
\ref{tab1}. Its $\gamma$-dependence is not monotonous and has a
nodal structure; the $z$-variation  is also non trivial. We have
remarked a strong dependence
 of $g(\gamma,z)$ relative to values of the $\varepsilon$
parameter smaller than $\sim10^{-4}$, in contrast to high stability of
corresponding eigenvalues.
However the corresponding  BS amplitude $\Phi$ and LF wave
function $\psi$, obtained from $g(\gamma,z)$ by the integrals
(\ref{bsint}) and (\ref{lfwf3a}), show the same strong stability
as the eigenvalues.

\begin{figure*}[ht!]
\centering
\includegraphics[width=8cm]{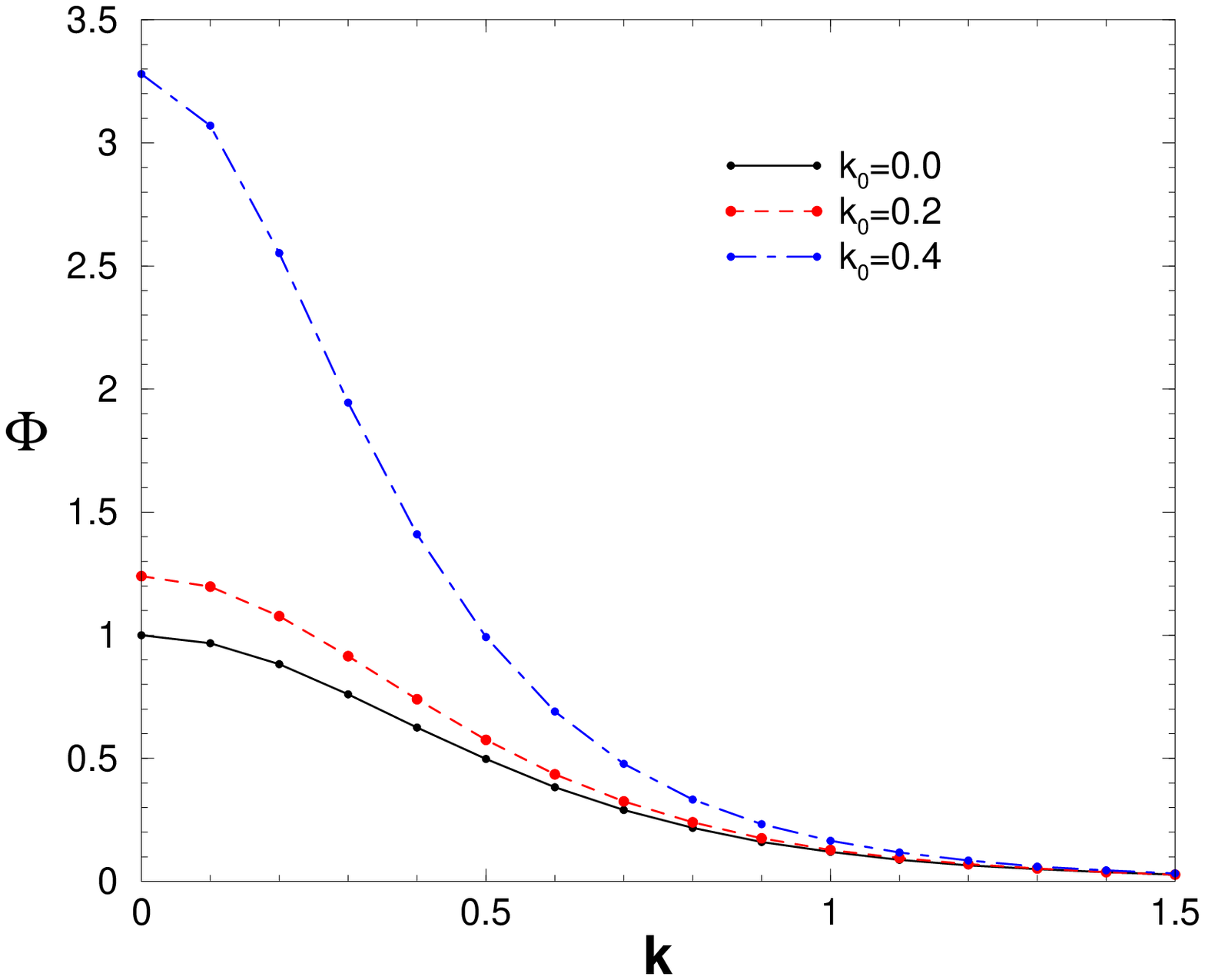}
\hspace{0.5cm}
\includegraphics[width=8cm]{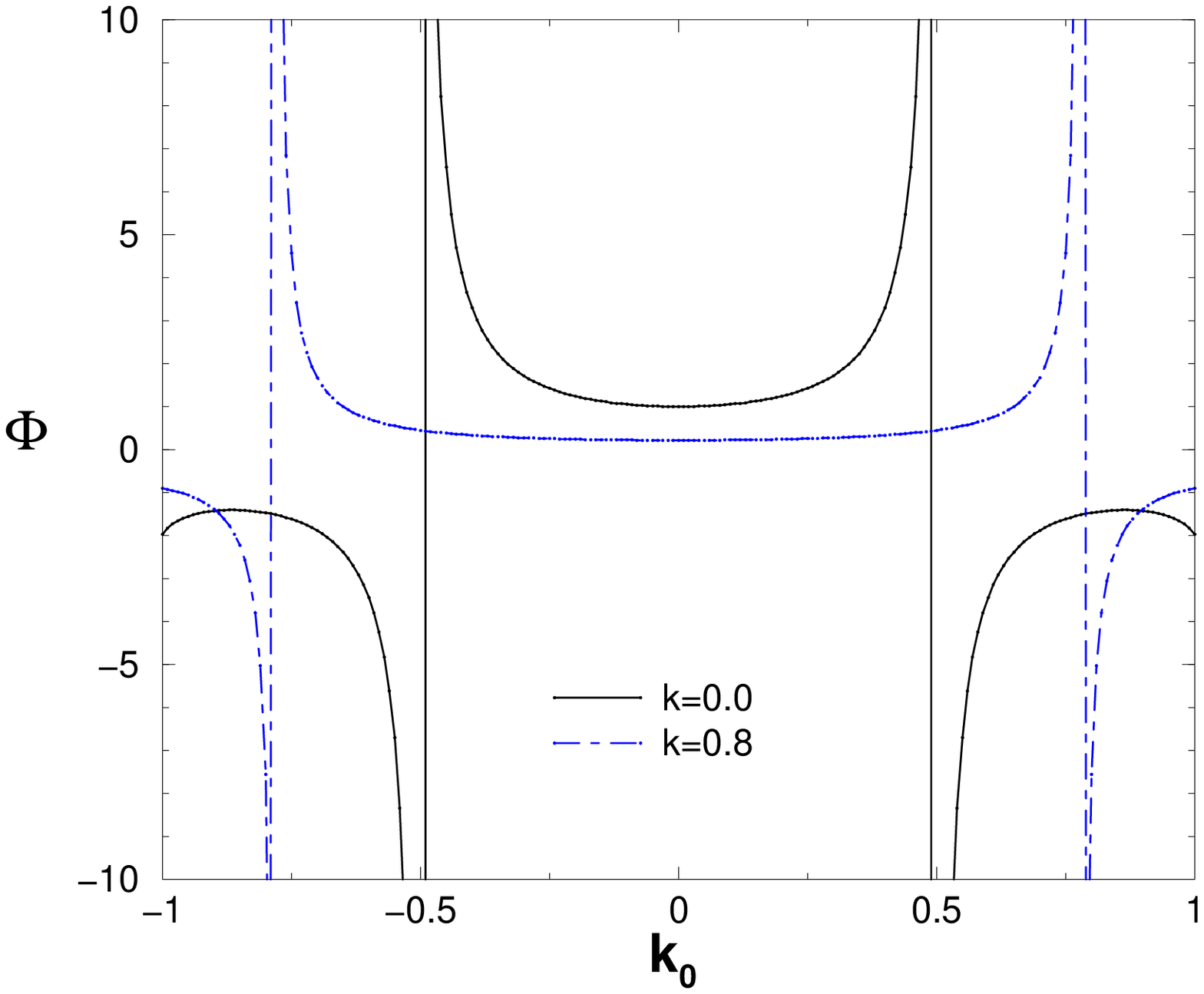}
\caption{BS amplitude  $\Phi(k_0,k)$ for $\mu=0.5$ and  $B=1.0$.
On left versus $k$ for fixed values of $k_0$ and
on right versus $k_0$ for a few fixed values of $k$.} \label{Phi_k}
\end{figure*}

The BS amplitude in Minkowski space in the rest frame $\vec{p}=0$ is shown
in fig. \ref{Phi_k}. The $k$-dependence is rather smooth but the
$k_0$-dependence, due to poles of the propagators in (\ref{bs}), exhibits
a singular behaviour at\\ $k_0=\pm \left(\varepsilon_k\pm
\frac{M}{2}\right)$, {\it i.e.} moving with $\vec{k}$ and $M$.

Note that our solution gives also the BS amplitude in Euclidean
space, by substituting in (\ref{bsint}) $k_0=ik_4$. The Euclidean
BS amplitude $\Phi_E(k_4,k)$ was found in this way in \cite{LC05}.
It is indistinguishable from the one obtained by a direct solution
of the Wick-rotated BS equation.

\begin{figure*}[ht!]
\centering
\includegraphics[width=8cm]{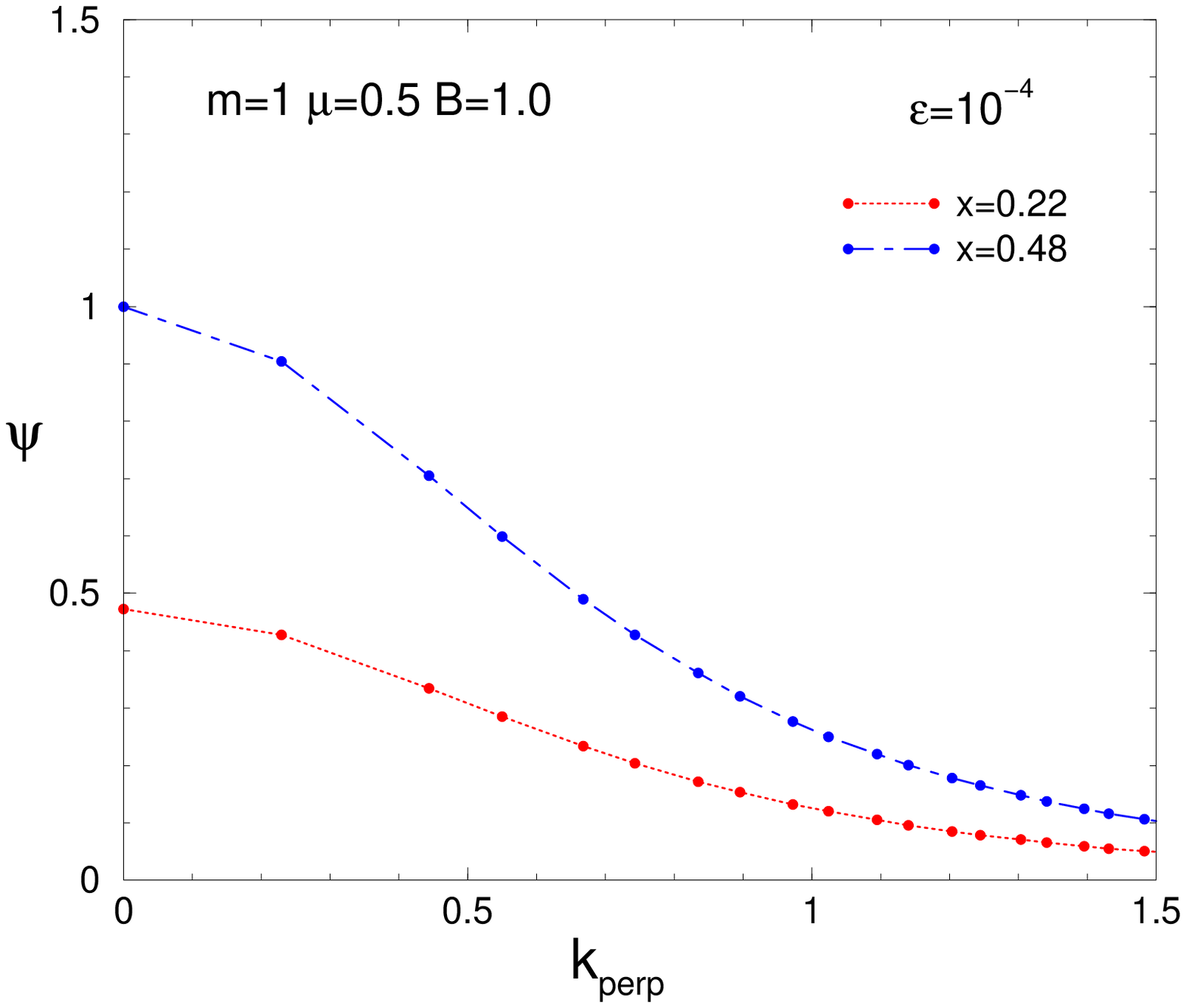}
\hspace{0.5cm}
\includegraphics[width=8cm]{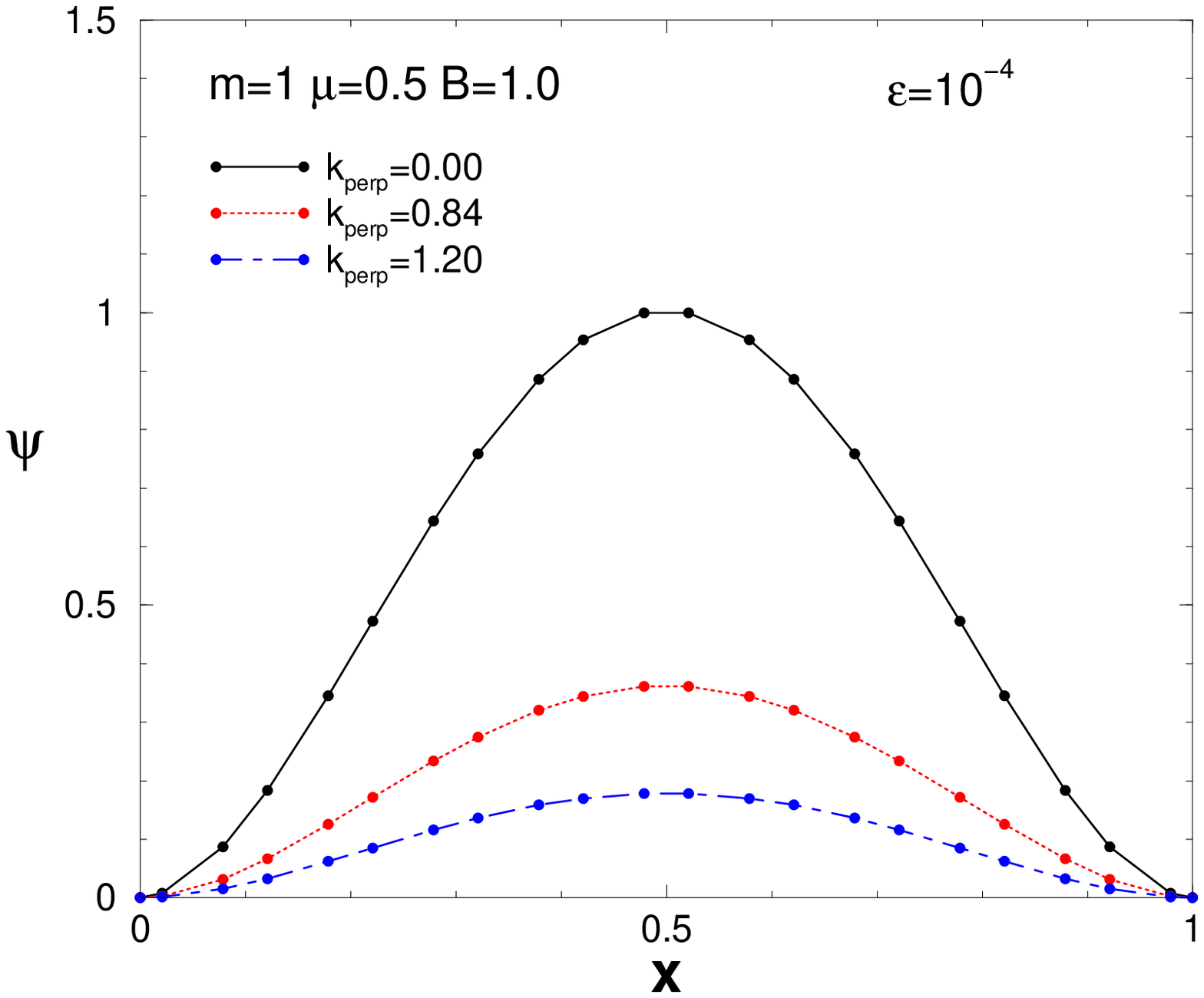}
\caption[*]{Wave function $\psi(k_{\perp},x)$ for $\mu=0.5$ and  $B=1.0$.
On left versus $k_{\perp}$ for fixed values of $x$ and
on right versus $x$ for a few fixed values of $k_{\perp}$.} \label{psi_k}
\end{figure*}

The corresponding LF wave function $\psi(k_{\perp},x)$ is shown in fig.
\ref{psi_k}. It is very similar to the LF wave functions displayed in ref.
\cite{mariane}, though they obey different dynamical equations. It has a
simpler structure than $g(\gamma,z)$ in both arguments.


\section{Conclusion}\label{concl}

We have developed a method for solving the Bethe-Salpeter equation
in the Minkowski space, {\it i.e.} without making use of the Wick
rotation.

The method is based on an integral transform of the original equation
which removes its singularities. It is motivated by the LF projection
(\ref{lfwf}) of the BS amplitude.

The transformed equation is formulated in terms of the weight
function of the Nakanishi integral representation \cite{nak63},
from which the original BS amplitude both in Minkowski and
Euclidean spaces as well as corresponding LF wave function can be
easily reconstructed.

The equation has been obtained for scalar particles and applied to
the ladder kernel. For zero-mass exchange the Wick-Cutkosky model
is derived. For massive exchange, numerical solutions have been
found. The binding energies are in full agreement with the
preceding results obtained in the Euclidean space. The singular BS
amplitude in Minkowski space has been displayed.

Calculation for the ladder exchange confirms the validity of our approach.
Our method can be used for an arbitrary kernel, given
by a Feynman graph. In the following paper \cite{ckII} it is applied  to
solve the BS equation with the cross-box kernel.

The method can be generalized to non-zero angular momentum and,
presumably, to the fermion case. A relation similar to
(\ref{lfwf}) between the BS amplitude and LF wave function for the
two-nucleon system is discussed in~\cite{cdkm,bbbd}.

\section*{Acknowledgements}
We are grateful to N.~Nakanishi for explaining to us conditions of
validity of his representation (\ref{bsint}), to V.I.~Korobov for
informing about the method of regularizing the kernel (\ref{reg})
and to M. Mangin-Brinet for providing the solutions of the ladder
BS equation in the Euclidean space. Numerical calculations were
performed at Institut du D\'e\-ve\-loppement et des Ressources en
Informatique Scienti\-fique (IDRIS) from  CNRS. One of the authors
(V.A.K.) is sincerely grateful for the warm hospitality of the
theory group at the Laboratoire de Physique Subatomique et
Cosmologie, Universit\'e Joseph Fourier, in Grenoble, where this
work was performed. This work is supported in part  by the RFBR
grant 05-02-17482-a (V.A.K.).

\appendix
\section{Derivation of equation (\ref{bsnew})}\label{deriv}
We substitute in equation (\ref{bs}) the BS amplitude in the form
(\ref{bsint}) and apply to both sides the transformation (\ref{lfwf}). For
the left-hand side this gives:
\begin{eqnarray}
\label{lfwf2}
&&\psi=\frac{1}{\pi}
\int_{-1}^1dz'\int_0^{\infty}d\gamma'\int_{-\infty}^{\infty}d\beta'
\\
&\times&
\frac{-i\left[1/4-(\omega\cdot k)^2/(\omega\cdot p)^2\right]g(\gamma',z')}
{\left[\gamma'+\kappa^2
-k^2-p\cdot k\; z'-\beta'\Bigl(z'+2\frac{(\omega\cdot k)}
{(\omega\cdot p)}\Bigr)-i\epsilon\right]^3},
\nonumber
\end{eqnarray}
where $\beta'=(\omega\cdot p)\beta$. As seen from (\ref{lfwf2}), $\psi$
depends on three scalar products $k^2$, $p\cdot k$ and $\frac{\omega\cdot
k}{\omega\cdot p}$. However, because of the relation
 $$ p\cdot
k=2\frac{\omega\cdot k} {\omega\cdot
p}\left(k^2-m^2+\frac{1}{4}M^2\right),
 $$
only two of them are independent. We use the LF variables (\ref{lfvar})
and express through them the scalar products:
\begin{eqnarray*}
k^2&=&m^2-\frac{k_{\perp}^2+m^2}{4x(1-x)},
\\
p\cdot k&=&\frac{1}{4}(1-2x)
\left(\frac{k_{\perp}^2+m^2}{x(1-x)}-M^2\right),
\\
\frac{\omega\cdot k}{\omega\cdot p}&=&x-\frac{1}{2}.
\end{eqnarray*}
By means of these relations, LF wave function (\ref{lfwf2}) depends on
$k_{\perp},x$.

It is convenient to introduce other notations:
$$
\gamma=k_{\perp}^2,\quad z=1-2x,\quad
\kappa^2=m^2-\frac{1}{4}M^2,
$$
so that:
\begin{eqnarray}
\label{kin1}
k^2&=&-\frac{(\gamma+z^2 m^2)}{1-z^2},
\nonumber\\
p\cdot k&=&\frac{z[\gamma+z^2 m^2+(1-z^2)\kappa^2]}{1-z^2},
\nonumber\\
\frac{\omega\cdot k}{\omega\cdot p}&=&-\frac{1}{2}z
\end{eqnarray}
and
\begin{equation}
\label{s}
s=\frac{4(\gamma+m^2)}{1-z^2}=\frac{k_{\perp}^2+m^2}{x(1-x)}.
\end{equation}

The integral (\ref{lfwf2}) over $\beta'$ is simply calculated by means of
the formula:
 $$ \int_{-\infty}^{\infty}\frac{d\beta}
{(\beta x -y-i\epsilon)^3}=\frac{i\pi}{y^2}\delta(x).
 $$
The result of transformation is given by eq. (\ref{lfwf3a}). In terms of
the variables $\gamma,z$ it reads:
\begin{equation}
\label{lfwf3b}
\psi(\gamma,z)=\frac{1}{8\sqrt{\pi}}\int_0^{\infty}\frac{(1-z^2)g(\gamma',z)d\gamma'}
{\Bigl[\gamma'+\gamma +z^2m^2+\kappa^2(1-z^2)\Bigr]^2}.
\end{equation}
Apart from the factor $(1-z^2)/(8\sqrt{\pi})$ (cancelled in the final
equation) it is the left-hand side of eq. (\ref{bsnew}). Substitution of
(\ref{G0}) and (\ref{bsint}) in right-hand side of (\ref{BST}) results in
the right-hand side of eq. (\ref{bsnew}).

The function $\psi(\gamma,z)$ in eq. (\ref{lfwf3b}) is the usual two-body
LF wave function. In terms of variables $k_{\perp},x$ it obtains the form
(\ref{lfwf3a}). The normalization integral reads:
\begin{eqnarray*}
N_2&=&\frac{1}{(2\pi)^3}\int \psi^2(k_{\perp},x)\;
 \frac{d^2 k_{\perp}dx}{2x(1-x)}
\\
&=& \frac{1}{(2\pi)^3}\int \psi^2(\gamma,z)\; \frac{\pi
d\gamma dz}{1-z^2}.
\end{eqnarray*}

We would like to emphasize the mathematical nature of the above
transformation. Namely,  the integral transformation (\ref{lfwf}), applied
to BS function (\ref{bsint}), obtains the form (\ref{lfwf2}). The function
$\psi$ there still depends on the variables  $k^2,p\cdot k$, like the
initial BS function, but it is not singular. However, the variables
$\gamma',z'$ in the integrand $g(\gamma',z')$ and $k^2,p\cdot k$ in $\psi$
run over different domains: $0\leq \gamma' < \infty$, $-1\leq z' \leq 1$,
whereas $-\infty < k^2 \leq 0$, $-\infty < p\cdot k < \infty$. By eqs.
(\ref{kin1}) we replace the variables  $k^2,p\cdot k$ by the new ones
$\gamma,z$, which, by construction, vary in the same domain as
$\gamma',z'$. In these variables eq. (\ref{lfwf2}) obtains the form
(\ref{lfwf3b}), where the functions $g(\gamma,z)$ and $\psi(\gamma,z)$
before and after integration are now defined in the same domain, like it
normally takes place in the integral equations. Therefore, in new
variables $\gamma,z$ we will find equation for $g(\gamma,z)$ in the domain
of its definition.

We separate the factor $(1-z^2)/(8\sqrt{\pi})$, {\it i.e.}
introduce $\tilde{\psi}$ related to $\psi$ as
$\psi(\gamma,z)=(1-z^2) \tilde{\psi}(\gamma,z)/(8\sqrt{\pi})$. That is:
\begin{equation}
\label{lfwf3}
\tilde{\psi}(\gamma,z)=\int_0^{\infty}\frac{g(\gamma',z)d\gamma'}
{\Bigl[\gamma'+\gamma +z^2m^2+\kappa^2(1-z^2)\Bigr]^2}.
\end{equation}
Eq. (\ref{bsnew}) can be rewritten for the function $\tilde{\psi}$:
\begin{equation}\label{psi}
\tilde{\psi}(\gamma,z)=
\int_0^{\infty}d\gamma'\int_{-1}^{1}dz'\;H(\gamma,z;\gamma',z')\tilde{\psi}(\gamma',z'),
\end{equation}
where $$ H(\gamma,z;\gamma',z')=\int_0^{\infty}d\gamma''
V(\gamma,z;\gamma'',z')h(\gamma'',\gamma',z') $$ and
$h(\gamma'',\gamma',z')$ is the kernel of the operator providing the
relation inverse to  eq. (\ref{lfwf3}), namely: $$
g(\gamma,z)=\int_0^{\infty}d\gamma'
h(\gamma,\gamma',z)\tilde{\psi}(\gamma',z). $$ Numerical calculation of
the kernel $h(\gamma,\gamma',z)$, then finding $H(\gamma,z;\gamma',z')$
and solving equation in the form (\ref{psi}) give the same results as for
(\ref{bsnew}).

\section{Derivation of the ladder LF equation (\ref{eq1})}
\label{LFeq} We take, for a moment,  $\omega=(1,0,0,-1)$, introduce the
variables $\vec{k'}_{\perp}=(k'_x,k'_y)$, $k'_{\pm}=k'_0\pm k'_z$ and
represent the integration volume $d^4k'$ in the right-hand side of eq.
(\ref{BST}) as
 $$d^4k'=\frac{1}{2}d^2k'_{\perp} dk'_+ dk'_-=
d^2k'_{\perp}dx'\;(\omega\cdot p) d\beta',
 $$
where $x'$ is defined in
(\ref{lfvar}) and $k'_-=\omega_-\beta'$ (with $\omega_-=2$). For arbitrary
$\omega$ eq. (\ref{BST}) obtains the form:
\begin{eqnarray}\label{BSTa}
&&\int d\beta\Phi(k+\beta\omega,p)=
\int d\beta G_0^{(12)}(k+\beta\omega,p)
\frac{d^2k'_{\perp}dx'}{(2\pi)^4}
\nonumber\\
&&\times
 iK(k+\beta\omega,k'+\beta'\omega,p)
\Phi(k'+\beta'\omega,p)(\omega\cdot p)
d\beta'
\end{eqnarray}

According to eq. (\ref{bs}), BS amplitude $\Phi(k,p)$ contains as a factor
the product of two free propagators (\ref{G0}). We separate the
propagators, {\it i.e.}, introduce the vertex function $\Gamma(k,p)$:
\begin{equation}
\label{Gam}
\Phi(k,p)=G^{(12)}_0(k,p)\Gamma(k,p)
\end{equation}
and substitute (\ref{Gam}) in the right-hand side of (\ref{BSTa}).

In order to derive LF equation (\ref{eq1}) from (\ref{BSTa}), we should,
integrating over $\beta'$, neglect the singularities of
$\Gamma(k'+\beta'\omega,p)$, {\it i.e.} deal with $\Gamma$ vs. $\beta'$ as
with a constant. That is, we will extract $\Gamma$ from integral over
$\beta'$ and introduce it back. This allows the following approximate
transformation of the right-hand side of (\ref{BSTa}) (for shortness we
show only the $\beta,\beta'$-dependence and don't show integration over
$d^2k'_{\perp}dx'$):
\begin{eqnarray*}
\int d\beta G_0^{(12)}(\beta) K(\beta,\beta')
 G_0^{(12)}(\beta') \Gamma(\beta') d\beta'\approx &&
\\
\Gamma\int d\beta G_0^{(12)}(\beta) K(\beta,\beta')
 G_0^{(12)}(\beta')d\beta'= &&
\\
\Gamma\frac{\int  G_0^{(12)}(\beta')d\beta'}
{\int  G_0^{(12)}(\beta')d\beta'}
\int d\beta G_0^{(12)}(\beta) K(\beta,\beta') G_0^{(12)}(\beta')d\beta'
\approx&&
\\
\frac{\int  G_0^{(12)}(\beta')\Gamma(\beta')d\beta'}
{\int  G_0^{(12)}(\beta')d\beta'}
\int d\beta G_0^{(12)}(\beta) K(\beta,\beta') G_0^{(12)}(\beta')d\beta'
=&&
\\
\frac{\int d\beta G_0^{(12)}(\beta) K(\beta,\beta')
 G_0^{(12)}(\beta')d\beta'}
{\int  G_0^{(12)}(\beta')d\beta'}
\int  \Phi(\beta')d\beta'\phantom{=}&&
\end{eqnarray*}
The integral $\int \Phi(\beta')d\beta'$ is understood as\\
$\int_{-\infty}^{\infty} \Phi(k'+\beta'\omega,p)d\beta'$ and it is related
by (\ref{lfwf}) to the LF wave function (as well as the left-hand side of
(\ref{BSTa})). With explicit expression (\ref{G0}) for $G_0^{(12)}$ we
find:
 $$
(\omega\cdot p)\int _{-\infty}^{\infty}d\beta
G_0^{(12)}(k+\beta\omega,p)=\frac{-\pi i}{x(1-x)(s-M^2)},
 $$
where $s$ is defined in (\ref{s}). With explicit expression (\ref{ladder})
for $K$ we obtain:
\begin{eqnarray*}
(\omega\cdot p)^2\int _{-\infty}^{\infty}d\beta
G_0^{(12)}(k+\beta\omega,p)
K(k+\beta\omega,k'+\beta'\omega,p)&&
\\
\times G_0^{(12)}(k'+\beta'\omega,p)d\beta'=&&
\\
\frac{-\pi i}{x(1-x)(s-M^2)}\;\left(-4m^2V_{LF}^{(L)}\right)\;
\frac{-\pi i}{x'(1-x')(s'-M^2)},&&
\end{eqnarray*}
where $V$ is the standard LF ladder kernel (see {\it e.g.} \cite{cdkm}):
\begin{eqnarray}\label{lfdlad}
&&V_{LF}^{(L)}(\vec{k}'_{\perp},x';\vec{k}_{\perp},x,M^2)=
\nonumber\\
&&-\frac{4\pi\alpha\theta(x'-x)}{(x'-x)(s_a-M^2)}
-\frac{4\pi\alpha\theta(x-x')}{(x-x')(s_b-M^2)},
\end{eqnarray}
and
\begin{eqnarray*}
s_a&=& \frac{\vec{k}^2_{\perp}+m^2}{x}
+\frac{(\vec{k'}_{\perp}-\vec{k}_{\perp})^2+\mu^2}{x'-x}+
\frac{\vec{k'}^2_{\perp}+m^2}{1-x'},
\\
s_b&=& \frac{\vec{k'}^2_{\perp}+m^2}{x'}
+\frac{(\vec{k'}_{\perp}-\vec{k}_{\perp})^2+\mu^2}{x-x'}+
\frac{\vec{k}^2_{\perp}+m^2}{1-x}.
\end{eqnarray*}
In this way we derive the equation (\ref{eq1}) with the kernel
(\ref{lfdlad}).

\section{Calculation of the kernel $V^{(L)}(\gamma,z;\gamma',z')$,
eq. (\ref{Kn})}
\label{calcI}
With the ladder kernel (\ref{ladder}) the integral  (\ref{I}) obtains the
form:
\begin{eqnarray*}
I(k,p)&=& \frac{-i 16\pi m^2\alpha}{(2\pi)^4} \int \frac{
d^4k'}{\Bigl[(k'-k)^2-\mu^2+i\epsilon\Bigr]}
\\
&\times&
\frac{1}{\Bigl[{k'}^2+p\cdot k'\;
z'-\kappa^2-\gamma'+i\epsilon\Bigr]^3},
\end{eqnarray*}
where we put $g^2=16\pi m^2\alpha$. Using the formula:
 $$ \frac{1}{a
b^3}=\int_0^1 \frac{3v^2 d v}{[a(1-v)+b v]^4}
 $$
and replacing $k'$ by new integration variable $q$ by the relation:
 $$ k'=q+(1-v)k-\frac{1}{2}v z'p, $$
we get
\begin{eqnarray}\label{I1}
I(k,p)&=&
\frac{-i 16\pi m^2\alpha}{(2\pi)^4}\int_0^1 3v^2 d v
\int\frac{d^4q}{\left[{q}^2+A(p,k)+i\epsilon\right]^4}
\nonumber\\
&=&\frac{\alpha m^2}{2\pi}\int_0^1 \frac{v^2 d v}{[A(p,k)+i\epsilon]^2}
\end{eqnarray}
with
\begin{eqnarray*}
A(p,k)&=&v(1-v)(k^2+p\cdot k\; z')-vm^2\\
&+&\frac{1}{4}v(1-v{z'}^2)M^2-(1-v)\mu^2-v\gamma'.
\end{eqnarray*}

Now we make in (\ref{I1}) the replacement $k\to k+\beta\omega$ and
substitute it in (\ref{V}). Scalar products $k^2,p\cdot k$ and also
$(\omega\cdot k)/(\omega\cdot p)$, which the kernel depends on, are
expressed through the variables $\gamma,z$ by eqs. (\ref{kin1}). Since
 $$
A(p,k+\beta\omega)=A(p,k)+\beta'v(1-v)(z'-z)
 $$
with $\beta'=(\omega\cdot p)\beta$, we obtain:
\begin{eqnarray}\label{rhs}
&&V^{(L)}(\gamma,z;\gamma',z')=\frac{-i\alpha m^2}{2\pi^2}
\nonumber\\
&&\times\int_0^1 v^2 d v
\int_{-\infty}^{\infty}\frac{d\beta'}
{\Bigl[A(p,k)+\beta'v(1-v)(z'-z)+i\epsilon\Bigr]^2}
\nonumber\\
&&\times
\frac{1}
{\left[k^2+p\cdot k-\kappa^2+(1-z)\beta'+i\epsilon\right]}
\nonumber\\
&&\times
\frac{1}
{\left[k^2-p\cdot k-\kappa^2-(1+z)\beta'+i\epsilon\right]}.
\end{eqnarray}
Both $z$ and $z'$ vary from $-1$ to 1. We consider two cases: ({\it
i})~$z' < z$ and ({\it ii})~$z< z'$. In the case  ({\it i}) the factor
$\Bigl(z'-z\Bigr)$ is negative and the (second order) pole in the variable
$\beta'$ of the factor $1/\Bigl[A+\beta'v(1-v)(z'-z)+i\epsilon\Bigr]^2$ in
(\ref{rhs}) is at the value $\beta'\sim \ldots  +i\epsilon$. We close the
contour in the lower half-plane, i.e, take the residue at the pole of the
first propagator:
 $$ \beta'=-\frac{k^2+p\cdot k-\kappa^2}{1-z}-i\epsilon.
 $$
This gives:
\[V^{(L)}(\gamma,z;\gamma',z')=W(\gamma,z;\gamma',z')\]
with
\begin{equation}
\label{V1}
W(\gamma,z;\gamma',z')=\frac{\alpha m^2(1-z)^2}
{2\pi\Bigl[\gamma +z^2 m^2+(1-z^2)\kappa^2\Bigr]}
\int_0^1\frac{v^2dv}{D^2}
\end{equation}
and
\begin{eqnarray*}
D&=&v(1-v)(1-z')\gamma+v(1-z)\gamma'\\
&+&v(1-z)(1-z')\Bigl[1+z(1-v) +v z'\Bigr]\kappa^2\\
&+&v\Bigl[(1-v)(1-z')z^2+v{z'}^2(1-z)\Bigr]m^2\\
&+&(1-v)(1-z)\mu^2.
\end{eqnarray*}

In the case ({\it ii}) the factor $(z'-z)$ is positive and the pole in the
variable $\beta'$ of the factor
$1/\Bigl[A+\beta'v(1-v)(z'-z)+i\epsilon\Bigr]^2$ is at the value
$\beta'\sim \ldots  -i\epsilon$. We close the contour in the upper
half-plane, i.e, take the residue at the pole of the second propagator:
 $$\beta'=\frac{k^2-p\cdot k-\kappa^2}{1+z}+i\epsilon.
 $$
This gives:
 $$ V^{(L)}(\gamma,z;\gamma',z')=W(\gamma,-z;\gamma',-z').
 $$
with $W$  defined in (\ref{V1}). The integral for $W$ is calculated
analytically. In this way, we obtain eqs. (\ref{Kn}), (\ref{W}) for the
ladder kernel in the equation (\ref{bsnew}).


\end{document}